\documentclass{article}
\usepackage{graphicx}
\usepackage{amstext,amssymb,amsbsy}
\usepackage{cite}

\author{
   Niels G. Gresnigt\\
  \texttt{niels.gresnigt@xjtlu.edu.cn}\\
  \small{Department of Mathematical Sciences}\\
  \small{Xi'an Jiaotong-Liverpool University}\\
  \small{111 Ren'ai Road, Suzhou HET, Jiangsu}\\
  \small{China 215123}\\
}

\title{Charge specific baryon mass relations with deformed $SU_q(3)$ flavor symmetry}
\begin{document}
\maketitle
\begin{abstract}

The quantum group $SU_q(3)=U_q(su(3))$ is taken as a baryon flavor symmetry. Accounting for electromagnetic contributions to baryons masses to zeroth order, new charge specific $q$-deformed octet and decuplet baryon mass formulas are obtained. These new mass relations have errors of only 0.02\% and 0.08\% respectively; a factor of 20 reduction compared to the standard Gell-Mann-Okubo mass formulas. A new relation between the octet and decuplet baryon masses that is accurate to 1.2\% is derived. An explicit formula for the Cabibbo angle, taken to be $\frac{\pi}{14}$, in terms of the deformation parameter $q$ and spin parity $J^P$ of the baryons is obtained.
\end{abstract}

\section{Introduction}
The aim of the present paper is to obtain octet and decuplet baryon mass relations of improved accuracy. This is achieved by first replacing the classical flavor symmetries by their quantum group counterparts (as in \cite{gavrilik1998quantum,gavrilik2001quantum,gavrilik1997quantum}) and then accounting for the electromagnetic contributions to baryons masses (as in \cite{morpurgo1992electromagnetic,morpurgo1992new,dillon2000miracle}).

Quantum groups (which are algebras rather than groups) provide a generalization of familiar symmetry concepts through the deformation of Lie groups and algebras. These deformations of classical Lie groups (algebras) into quantum groups (algebras) extends the domain of classical group theory. Such quantum groups are deformations on Hopf algebras and depend on a deformation parameter $q$ with the value $q=1$ returning the undeformed universal enveloping algebra. First formalized by Jimbo \cite{jimbo1985aq} and Drinfeld \cite{drinfeld1985soviet} as a class of Hopf algebras, quantum groups have found many applications in theoretical physics, see \cite{finkelstein2001q,Finkelstein2012,Steinacker1998,Sternheimer2007,Majid1994,Lukierski2003,castellani1996quantum} and references therein. 

The applications of quantum groups to hadron phenomenology, and in particular as flavor symmetries have been explored in \cite{gavrilik2004quantum,gavrilik2001quantum,gavrilik1998quantum,gavrilik1997quantum,carcamo2005gell}.
The standard Gell-Mann-Okubo mass formula \cite{gell1961eightfold,okubo1962note}, a result of $SU(3)$ flavor symmetry, is
\begin{eqnarray}
M=\alpha_0+\alpha_1 Y+\alpha_2 \left[ I(I+1)-\frac{1}{4} Y^2\right], 
\end{eqnarray}
where $M$ is the mass of a hadron within a specific multiplet, $Y$ and $I$ are the hypercharge and isospin respectively, and $\alpha_0,\alpha_1, \alpha_2$ are free parameters. By eliminating the free parameters $\alpha_0,\alpha_1, \alpha_2$, one obtains mass relations between the different baryons within a given multiplet. 
For the case of octet baryons $(1/2)^+$ one obtains the standard relation
\begin{eqnarray}\label{GMOoctet}
N+\Xi=\frac{3}{2}\Lambda+\frac{1}{2}\Sigma,
\end{eqnarray} 
whereas for the decuplet baryons $(3/2)^+$ one obtains the equal spacing rule
\begin{eqnarray}\label{GMOdecuplet}
\Delta-\Sigma^*=\Sigma^*-\Xi^*=\Xi^*-\Omega
\end{eqnarray}
These relations hold to first order in flavor symmetry breaking only. Using the most recent data formula (\ref{GMOoctet}) is accurate to about 0.6\%. The equal spacing rule (\ref{GMOdecuplet}) is less accurate ($\Delta-\Sigma^*=-152.6$MeV, $\Sigma^*-\Xi^*=-148.8$MeV, $\Xi^*-\Omega=-139.1$MeV), however a modified relation $\Omega-\Delta=3(\Xi^*-\Sigma^*)$ due to Okubo \cite{okubo1963some} is accurate to about 1.4\%.

Adopting instead $q$-deformed flavor symmetries obtained by replacing the unitary flavor symmetry $SU(3)$ by its quantum group counterpart $SU_q(3)\equiv U_q(su(3))$, Gavrilik was able to derive deformed mass relations for octet and decuplet baryons of increased accuracy \cite{gavrilik1997quantum,gavrilik2001quantum,gavrilik2004quantum}. By fixing a definite value for the deformation parameter $q$ (through fitting the data), the modified baryon mass sum rules are accurate to an impressive 0.06\% for the octet baryons and 0.32\% for the decuplet baryons. 

The masses used in both the standard as well as the $q$-deformed mass relations are the averages of the isospin multiplets (isoplets). We argue here that at the level of accuracy of the $q$-deformed mass relations, the mass splittings within isoplets become significant and should be accounted for. For example $\Sigma^- - \Sigma^+=8$MeV, $\Xi^{*-}-\Xi^{*0}=3.2$MeV and $\Sigma^{*-}-\Sigma^{*+}=4.4$MeV, representing about $\sim 0.2-0.4\%$ of the average isoplet mass. The impressive accuracy of the deformed mass relations lose their significance when these mass splittings (due to electromagnetic contributions to the masses of baryons) are ignored because the errors in these relations are less than the variation of masses within the isoplets.

Our starting point is the $q$-deformed mass relations derived in \cite{gavrilik1998quantum,gavrilik2001quantum,gavrilik1997quantum}. We then make use of the QCD general parametrization scheme of Morpurgo  \cite{morpurgo1989field} in which the electromagnetic contributions (to zeroth order flavor symmetry breaking) to octet and decuplet baryon masses are expressed in terms of four parameters. Applying this parametrization to the deformed mass relations allows us to derive relations that have equal electromagnetic contributions to mass on both sides. The resulting charge specific $q$-deformed baryon mass relations are exceptionally accurate, holding to within 0.02\% and 0.08\% for octet $(1/2)^+$ and decuplet $(3/2)^+$ baryons respectively. 

The values of the deformation parameter $q$ that give these exceptionally accurate octet and decuplet baryons, together with the assertion made in \cite{gavrilik2004quantum,gavrilik2001can} that there should be a direct connection between $q$ and the Cabibbo angle conspire to suggest a particularly simple formula for the Cabibbo angle in terms of $q$ and spin parity $J^P$.

In section \ref{qgsection} a very brief overview of quantum groups is given and their application as flavor symmetries discussed. Section \ref{emsection} focuses on electromagnetic mass splittings within isoplets and how to account for these electromagnetic contributions to baryon masses before charge specific and $q$-deformed mass relations for octet and decuplet baryons are presented in section \ref{newsection}. Finally in section \ref{cabibbosection} we look at the connection between $q$ and the Cabibbo angle.

\section{$q$-Deformed baryon mass relations}\label{qgsection}

\subsection{Quantum groups and algebras}
Only a very brief introduction to the relevant aspects of quantum groups is provided here. The literature on quantum groups and algebras is extensive. For an excellent introduction the reader is directed to \cite{jaganathan2001some}. The manifold applications of quantum groups to physics are discussed in \cite{castellani1996quantum}.


The quantum (enveloping) algebra $SU_q(n)\equiv U_q(su(n))$ corresponding to a one-parameter deformation of the universal enveloping algebra of $su(n)$, is a Hopf algebra with unit $\mathbf{1}$ and generators $H_i$, $X_i^{\pm}$, $i=1,2,...,n-1$, defined through the commutation relations in the Cartan-Chevalley basis as 
\begin{eqnarray}
\left[ H_i, H_j\right] &=&0\\
\left[ H_i, X_j^{\pm}\right] &=&a_{ij}X_j^{\pm}\\
\left[ X_i^+, X_j^-\right]&=&\delta_{ij}[H_i]_q\equiv\delta_{ij}\frac{q^{H_i}-q^{-H_i}}{q-q^{-1}}\label{quantalg3},
\end{eqnarray}
together with the quadratic and cubic deformed $q$-Serre relations
\begin{eqnarray}
\left[ X_i^{\pm}, X_j^{\pm}\right]=0,\;\; j\neq i\pm 1,\;\; 1\leq i,j\leq n-1
\end{eqnarray}
and
\begin{eqnarray}\label{cubicserre}
(X_i^{\pm})^2 X_j^{\pm}-[2]_q X_i^{\pm}X_j^{\pm}X_i^{\pm}+X_j^{\pm}(X_i^{\pm})^2=0,\;\; j=i\pm 1,\;\; 1\leq i,j\leq n-1
\end{eqnarray}
respectively \cite{jimbo1985aq,quesne1992complementarity}. Here $a_{ij}$ is an element of the Cartan matrix
 \begin{displaymath}
   a_{ij} = \left\{
     \begin{array}{lr}
       2 & j=i\\
       -1 & j=i\pm 1\\
       0 & \textrm{otherwise}.
     \end{array}
   \right.
\end{displaymath} 
The $q$-number
\begin{eqnarray}
[N]_q=\frac{q^N-q^{-N}}{q-q^{-1}}
\end{eqnarray}
is defined for both operators (as in equation (\ref{quantalg3})) and real numbers\footnote{In this paper we will only have to deal with integer values of $N$. The definition however holds for real numbers.} (as in equation (\ref{cubicserre})). The definition of the algebra is completed by the Hermiticity properties
\begin{eqnarray}
(H_i)^{\dagger}=H_i,\qquad (X_i^{\pm})^{\dagger}=X_i^{\mp}.
\end{eqnarray}

The quantum algebra $SU_q(n)$ has the structure of a Hopf algebra admitting a coproduct, counit and antipode. These are not used here and so we do not define them (see \cite{quesne1992complementarity}). In the limit $q=1$ the above relations approach the relations for the universal enveloping algebra $U(su(n))$ but for general $q$ they represent a deformation of the universal enveloping algebra of $su(n)$.

The representation theory of quantum groups in general differs from that of classical groups, especially when $q$ is taken to be complex. For complex $q$, the universal enveloping algebra becomes complex, admitting non-unitary representations. A special case occurs when $q$ is a root of unity. In this case there exist (finite dimensional) unitary representations. Even in this case however, the $q$-tensor product of unitary irreps need no longer be unitary and not all irreps are completely reducible \cite{castellani1996quantum,steinacker1998finite}. 

\subsection{Quantum groups as flavor symmetries}

In \cite{gavrilik1998quantum,gavrilik1997quantum,gavrilik2001quantum,gavrilik2004quantum}, the $q$-analogues $SU_q(n)\equiv U_q(su(n))$ instead of the groups $SU(n)$ are taken as hadronic flavor symmetries and improved baryon mass relations are derived. The basic approach of the construction is the representation theory of $U_q(su(n))$ \cite{gavrilik1995representations,gavrilik2004quantum}. $q$-Deformed mass relations are computed from the expectation value of the mass operator which is defined in terms of the generators of the dynamical algebras (quantum groups) $U_q(u(n+1))$ or $U_q(u(n,1))$. The expectation values are computed from the matrix elements of these generators. Utilizing the $q$-algebras $U_q(u(n+1))$ or $U_q(u(n,1))$ of dynamical symmetry, breaking of $n$-flavor symmetries up to exact isospin symmetry $SU_q(2)$ are realized and the $q$-analogues of mass sum rules for baryon multiplets are derived. 

The general procedure (see \cite{gavrilik1995representations,gavrilik1997quantum,gavrilik1998quantum} for details) is as follows:
\begin{enumerate}
\item Use the Gelfand-Tsetlin basis vectors for baryon states of $n$-flavor $U_q(u(n))$ embedded into dynamical $U_q(u(n+1))$,
\vspace{-0.25cm}
\item Construct a mass operator $\hat{M}_n$ invariant under the isospin+hypercharge deformed $U_q(u(2))$ from the generators of the dynamical algebra $U_q(u(n+1))$,
\vspace{-0.25cm}
\item Calculate expressions for masses $M_{B_i}=<B_i\vert \hat{M}_n\vert B_i>$ (with $\vert B_i>$ suitably defined) involving $M_0$ and flavor symmetry breaking parameters $\alpha$ and $\beta$ as well as the deformation parameter $q$,
\vspace{-0.25cm}
\item Exclude the undetermined parameters (except $q$) from final expressions for $M_{B_i}$ to obtain $q$-deformed baryon mass relations.
\end{enumerate}
 
\subsection{Octet baryons}

With the restriction that $q=q_n=e^{i\pi/n}$ for integer $n$, the $q$-deformed mass relation for octet baryons obtained in \cite{gavrilik1997quantum} is
\begin{eqnarray}\label{generaloctet}
N+\frac{1}{[2]_{q_n}-1}\Xi=\frac{[3]_{q_n}}{[2]_{q_n}}\Lambda+\left( \frac{[2]_{q_n}}{[2]_{q_n}-1}-\frac{[3]_{q_n}}{[2]_{q_n}}\right) \Sigma.
\end{eqnarray}

The deformed mass relations depend on the deformation parameter $q$ and the set of integers $n$ produces an infinite set of mass relations. The value on $n$ that leads to the closest agreement with  experimental data \cite{olive2014review} is $n=7$ where it is the average mass of isoplets that has been used\footnote{For a table where the accuracy of the mass relations is evaluated for different values of $n$, see Gavrilik \cite{gavrilik1997quantum}.}. When $q=e^{i\pi/7}$, $[3]_{q_7}=\frac{[2]_{q_7}}{[2]_{q_7}-1}$ and the mass relation (\ref{generaloctet}) simplifies to
\begin{eqnarray}\label{defoctet}
N+\frac{\Xi}{[2]_{q_7}-1}=\frac{\Lambda}{[2]_{q_7}-1}+\Sigma,\;\;q_7=e^{i\pi/7},
\end{eqnarray}
which is accurate to $0.06\%$! this represents a very significant tenfold increase in accuracy compared to the GMO formula (\ref{GMOoctet}) which is accurate to about 0.6\%. The accuracy to which the formula (\ref{defoctet}) holds is surprising given that the masses within an isoplet can differ by around $\sim 0.5\%$ whereas the masses used above are the averages of the isoplets.

\subsection{Decuplet baryons}

For the decuplet baryons, the equal spacing rule (\ref{GMOdecuplet}) is deformed to\cite{gavrilik1995representations}
\begin{eqnarray}\label{defdecup}
(\Sigma^*-\Delta+\Omega-\Xi^*)=[2]_q(\Xi^* -\Sigma^*).
\end{eqnarray}

Rearranging, (\ref{defdecup}) may be written as
\begin{eqnarray}
\Omega-\Delta=(1+[2]_q)(\Xi^*-\Sigma^*),
\end{eqnarray}
which is reminiscent of the relations obtained by Okubo \cite{okubo1963some}
\begin{eqnarray}
\Omega-\Delta= 3(\Xi^*-\Sigma^*),
\end{eqnarray}
that, unlike the equal spacing rule holds for second order flavor symmetry breaking.

Taking $q=e^{i\pi/n}$ one finds excellent agreement with data for $n=14$. We note however that $n=14$ does not provide the best fit to the experimental data. Solving for integer $n$ suggests $n=16$ should provide the best fit. Experimental uncertainties of the masses are however ignored in this analysis and a range of values for $n$ provide an excellent fit to the data. The choice of $n=14$ in \cite{gavrilik1997quantum, gavrilik2001quantum} is motivated by the observation that $[2]_{q_{14}}$ is readily related to $[2]_{q_{7}}$ (the best fit for the $q$-deformed octet formula)
\begin{eqnarray}
([2]_{q_{14}})^2= q_{14}^2+2+q_{14}^{-2}=q_7+q^{-1}_{7}+2=[2]_{q_7}+2.
\end{eqnarray}
Solving both (\ref{defoctet}) and (\ref{defdecup}) for $[2]_{q_n}$ and using the above relation, one obtains a new octet-decuplet mass relation \cite{gavrilik1997quantum}
\begin{eqnarray}\label{defoctdec}
\frac{\Omega-\Xi^*+\Sigma^*-\Delta}{\Xi^*-\Sigma^*}=\left( 3+\frac{\Xi-\Lambda}{\Sigma-N}\right) ^{1/2}.
\end{eqnarray}
Using the latest data and averaging the isoplet masses, this relation holds within $\sim 1.5\%$.

\section{Electromagnetic contributions}\label{emsection}

The masses within a given isoplet of a baryon multiplet differ by a few tenths of a percent. This mass splitting is the result of the electromagnetic contributions to baryon masses. The electromagnetic contributions to baryon masses may be determined within the QCD general parametrization scheme in the spin-flavor space considered by Morpurgo \cite{morpurgo1989field}. To zeroth order symmetry breaking the electromagnetic contributions to the octet baryon masses are given in terms of four parameters as \cite{morpurgo1992new}
\begin{eqnarray*}
\delta_0p&=&\mu+\frac{5}{9}\nu+\eta+\rho\qquad
\delta_0n=\frac{2}{3}\mu\\
\delta_0\Lambda&=&\frac{2}{3}\mu+\frac{1}{9}\nu\qquad
\delta_0\Sigma^+=\mu+\frac{5}{9}\nu+\eta+\rho\\
\delta_0\Sigma^-&=&\frac{1}{3}\mu+\frac{1}{9}\nu+\eta+\frac{1}{3}\rho\qquad
\delta_0\Sigma^0=\frac{2}{3}\mu+\frac{1}{3}\nu\\
\delta_0\Xi^0&=&\frac{2}{3}\mu \qquad
\delta_0\Xi^-=\frac{1}{3}\mu+\frac{1}{9}\nu+\eta+\frac{1}{3}\rho.
\end{eqnarray*}
It is easily checked that
\begin{eqnarray}
\delta_0 p+\delta_0\Xi^0=\frac{3}{2}\delta_0\Lambda^0+\frac{1}{2}(2\delta_0\Sigma^+-\delta_0\Sigma^0).
\end{eqnarray}
For the octet baryons, accounting fot the electromagnetic mass contributions the standard Gell-Mann-Okubo formula (\ref{GMOoctet}) becomes \cite{dillon2000miracle}
\begin{eqnarray}\label{EMoctet}
\frac{1}{2}(p+\Xi^0)+T=\frac{1}{4}(3\Lambda+2\Sigma^+-\Sigma^0),
\end{eqnarray}
where $T=\Xi^{*-}-\frac{1}{2}(\Omega^-+\Sigma^{*-})=5.18$MeV is a decuplet correction. This equation has equal electromagnetic mass contributions on both sides of the equation (the electromagnetic contribution of the decuplet correction is zero, $\delta_0T=0$). This equation holds to within $\sim 0.13$\%, a significant improvement over the standard GMO octet relation.

Applying the same parametrization procedure to the decuplet baryons, one finds that\footnote{It is worth noting that the second order flavor symmetry breaking affects only baryons with a strangeness of 2 or 3, namely $\Xi,\Xi^*$ and $\Omega$.}
\begin{eqnarray}
\delta_0\Omega^-=\delta_0\Sigma^{-*}=\delta_0\Xi^{-*}=\delta_0\Delta^-.
\end{eqnarray}
Consequently, the Okubo relation becomes the charge specific decuplet mass relation
\begin{eqnarray}
\Omega^--\Delta^-=3(\Xi^{*-}-\Sigma^{*-}),
\end{eqnarray}
which is accurate to $\sim 0.67$\%, a factor of two improvement over Okubo's relation. 

\section{Charge specific q-deformed baryon mass relations}\label{newsection}

As discussed in the introduction, the impressive accuracy of the $q$-deformed baryon mass relations means that mass splitting within isoplets can no longer be ignored. To account for these mass splitting the specific charges or baryons must be inserted into the mass relations in such a way that the electromagnetic contribution to masses is the same on both sides of the equation. In this section we apply the work discussed in section \ref{emsection} to the $q$-deformed octet and decuplet mass relations (\ref{defoctet}) and (\ref{defdecup}). What we find is remarkably accurate charge specific $q$-deformed mass relations for both octet and decuplet baryons.

\subsection{New octet baryon mass relation}

Our starting point is the $q$-deformed octet baryon mass relations, eqn. (\ref{generaloctet}). We now apply the parametrization of Morpurgo\cite{morpurgo1992new} in order to rewrite the deformed mass relations in such a way that both sides have equal electromagnetic mass contributions. Although somewhat more tedious than the case described in the previous section, one may check that

\begin{eqnarray}\label{newoctetbaryongeneral}
p+\frac{2[3]_q}{3[2]_q}\Xi^0+\left( \frac{1}{[2]_q-1}-\frac{2[3]_q}{3[2]_q}\right)\Xi^-=\frac{[3]_q}{[2]_q}\Lambda-\frac{[3]_q}{3[2]_q}\Sigma^0+\left( \frac{1}{[2]_q-1}-\frac{2[3]_q}{3[2]_q}\right)\Sigma^-+\Sigma^+
\end{eqnarray}
is unaffected by the electromagnetic corrections (to zeroth order in flavor breaking). 

Substituting the experimental masses for the baryons, we may determine the value of  $q=q_n=e^{i\pi/n}$ for integer $n$ that minimizes the error in the above mass relation. One finds that $n=7$ continues to provide the best fit to the experimental data. For this value of $q$, equation (\ref{newoctetbaryongeneral}) simplifies to
\begin{eqnarray}\label{newoctet}
p+\frac{(2\Xi^0+\Xi^-)}{3([2]_{q_7}-1)}=\frac{\Lambda}{[2]_{q_7}-1}+\frac{(\Sigma^- -\Sigma^0)}{3([2]_{q_7}-1)}+\Sigma^+,
\end{eqnarray}
which has an error of only $0.02\%$, a threefold reduction in error compared to the $q$-deformed mass relation (\ref{generaloctet}) which ignores electromagnetic contributions to mass. 

A comparison of the different octet baryons mass relations and their accuracy is presented in Table \ref{table1},
\begin{table}[h]
\small
\begin{center}\label{table1}
  \begin{tabular}{| l || l | c | c| c |}
    \hline
    & GMO & Charge specific \cite{morpurgo1992new}& $q$-deformed \cite{gavrilik1997quantum} & Eqn.(\ref{newoctet})\\ \hline
   \hline
  LHS (MeV)& 2257.204 & 2263.532 & 2585.871 & 2582.012 \\ \hline
  RHS (MeV) & 2270.102 & 2266.574 & 2584.453 & 2582.598 \\ \hline
  Error (\%) & 0.57 & 0.13 & 0.06 & 0.02 \\ \hline 
  $\vert LHS-RHS \vert$& 12.9 & 3.1 & 1.4 & 0.6 \\ 
    \hline
  \end{tabular}
  \caption{\footnotesize Table comparing the accuracy of the standard Gell-Mann-Okubo (GMO) octet mass relations with several generalization including the charge specific octet generalization that takes into account the electromagnetic mass splitting within isoplets, the $q$-deformed octet mass relation, and the new octet formula presented in this paper that takes into account both $q$-deformation and electromagnetic contributions to baryon masses. Data was obtained from \cite{olive2014review}.}
\end{center}
\end{table}
where we have used:
\begin{eqnarray*}
\textrm{GMO}&:& N+\Xi=\frac{3}{2}\Lambda+\frac{1}{2}\Sigma\\
\textrm{Charge specific}&:& p+\Xi^0+2T=\frac{3}{2}\Lambda+\frac{1}{2}(2\Sigma^+-\Sigma^0),\;\;T=5.2MeV\\
\textrm{$q$-deformed}&:& N+\frac{\Xi}{[2]_{q_7}-1}=\frac{\Lambda}{[2]_{q_7}-1}+\Sigma,\;\;q_7=e^{i\pi/7}\\
\textrm{Eqn.(\ref{newoctet})}&:& p+\frac{(2\Xi^0+\Xi^-)}{3([2]_{q_7}-1)}=\frac{\Lambda}{[2]_{q_7}-1}+\frac{(\Sigma^- -\Sigma^0)}{3([2]_{q_7}-1)}+\Sigma^+. 
\end{eqnarray*}

\subsection{New decuplet baryon mass relation}
Similarly to the octet case, accounting for the electromagnetic mass contribution to the $SU_q(3)$ deformed decuplet mass relation (\ref{defdecup}) leads to the new decuplet formula
\begin{eqnarray}\label{newdecuplet}
\Omega^--\Delta^-=([2]_q+1)(\Xi^{*-}-\Sigma^{*-}).
\end{eqnarray}
In \cite{gavrilik1997quantum} it was suggested that choosing $q=e^{i\pi/14}$ provides a good fit to data. Although this remains true in the present case, there are other values of $q$ for which the error is smaller. In particular, we consider the case where $q=e^{i\pi/21}$. Although this is not the absolute best choice for $q$ \footnote{Putting the experimental data into \ref{newoctet}, substituting $q=e^{i\pi/n}$ gives $n=22$ (for integer $n$). The error that formula \ref{GMOdecuplet} give when $n=22$ is just 0.01\%!}, because 21 is a multiple of 7, it allows for an elegant new relation between octet and decuplet baryons (this is also the reason why $n=14$ is chosen in \cite{gavrilik1997quantum,gavrilik2001quantum}). Taking $n=21$ gives LHS=440.45MeV and RHS=440.10MeV, an error of only 0.08\%. The different decuplet formulas and their accuracy are summarized in Table \ref{table2}, 
\begin{table}[h]
\small
\begin{center}\label{table2}
  \begin{tabular}{| l || c | c | c| c |}
    \hline
    & Okubo \cite{okubo1962note} & Charge specific \cite{morpurgo1992new} &$q$-deformed \cite{gavrilik1997quantum}  & Eqn.(\ref{newdecuplet})\\ \hline
   \hline
  LHS (MeV)& 440.45 & 440.45 & 440.45 & 440.45 \\ \hline
  RHS (MeV)& 446.49 & 443.40 & 439.05 & 440.15 \\ \hline
  Error (\%) & 1.4 & 0.67 & 0.32 & 0.07 \\ \hline 
  $\vert LHS-RHS \vert$ & 6.0 & 3.0 & 1.4 & 0.3 \\ 
    \hline
  \end{tabular}
  \caption{\footnotesize Table comparing the accuracy of the standard Okubo decuplet mass relation with several generalization including the charge specific decuplet generalization that takes into account the electromagnetic mass splitting within isoplets, the $q$-deformed decuplet mass relation, and the new decuplet formula presented in this paper that takes into account both $q$-deformation and electromagnetic contributions to baryon masses. Data was obtained from \cite{olive2014review}.}
\end{center}
\end{table}
where we have used:
\begin{eqnarray*}
\textrm{Okubo}&:& \Omega-\Delta=3(\Xi^*-\Sigma^*)\\
\textrm{Charge specific}&:& \Omega^--\Delta^-=3(\Xi^{*-}-\Sigma^{*-})\\
\textrm{$q$-deformed}&:& \Omega-\Delta=([2]_{q_{14}}+1)(\Xi^*-\Sigma^*),\;\;q_{14}=e^{i\pi/14}\\
\textrm{Eqn.(\ref{newdecuplet})}&:& \Omega^--\Delta^-=([2]_{q_{21}}+1)(\Xi^{*-}-\Sigma^{*-}),\;\;q_{21}=e^{i\pi/21}.
\end{eqnarray*}

\subsection{Octet-decuplet mass relation}
Both formulas (\ref{newoctet}) and (\ref{newdecuplet}) provide excellent fits to experimental data. The octet formula is valid only for $q=e^{i\pi/7}$ whereas the decuplet formula is valid for all any $q_n=e^{i\pi/n}$, where $n$ is an integer. In particular we chose $n=21$ as it provides (with the exception of $n=22$) the best fit to data. Because 21 is a multiple of 7, we can find a relation between octet and decuplet baryon masses by solving both mass relations for  $[2]_{q_7}$ and $[2]_{q_{21}}$ respectively and using the fact that $[2]_{q_{21}}^3= (q_{21}+q_{21}^{-1})^3=[2]_{q_7}+3[2]_{q_{21}}$ to obtain

\begin{eqnarray}\label{newoctdec}
\left(\frac{\Omega^--\Delta^-}{\Xi^{*-}-\Sigma^{*-}}-1\right) ^3-3\left( \frac{\Omega^--\Delta^-}{\Xi^{*-}-\Sigma^{*-}}-1\right)=\left( 1+\frac{\Lambda+\frac{1}{3}\left(\Sigma^--\Sigma^0-2\Xi^0-\Xi^-\right)}{\left(p-\Sigma^+\right)}\right). 
\end{eqnarray}
This formula has an error of $\sim 1.2\%$; a slight improvement over the formula in \cite{gavrilik1997quantum,gavrilik2001quantum} with the choice $n=14$ (and error $\sim 1.5\%$).

\section{The Cabibbo angle as a function of $q$ and $J^P$}\label{cabibbosection}

It is asserted in \cite{gavrilik2004quantum,gavrilik2001can} that there should be a direct connection between the deformation parameter $q$ and the Cabibbo angle $\theta_C$. 

With the choice of $\theta_{\mathbf{8}}=\frac{\pi}{7}$, such that $q=e^{i\theta_{\mathbf{8}}}$ for octet baryons and  $\theta_{\mathbf{10}}=\frac{\pi}{14}$, such that $q=e^{i\theta_{\mathbf{10}}}$ for decuplet baryons, the suggested explicit form between these angles (which are related to $q$) and the Cabibbo angle is
\begin{eqnarray}\label{cabibbo}
\theta_{\mathbf{10}}=\theta_C,\qquad \theta_{\mathbf{8}}=2\theta_C.
\end{eqnarray}
The Cabibbo angle here takes the exact value $\theta_C=\frac{\pi}{14}$.

These choices for $\theta_{\mathbf{8}}$ and $\theta_{\mathbf{10}}$ are a result of fitting data to the $q$-deformed octet and decuplet mass sum rules in \cite{gavrilik1997quantum,gavrilik2001quantum}. Once the electromagnetic contributions to mass are taken into account however, the choice $q=e^{i\theta_{\mathbf{10}}}=e^{i\pi/21}$ is superior, providing a significantly better fit to data than $\theta_{\mathbf{10}}= \frac{\pi}{14}$. For these improved formulas, the explicit form (\ref{cabibbo}) no longer holds true. Instead we write
\begin{eqnarray}
\theta_C=\frac{1}{2}\theta_{\mathbf{8}}=\frac{3}{2}\theta_{\mathbf{10}}=\frac{\pi}{14},
\end{eqnarray}
Making the observation that for octet baryons $J^p=(1/2)^+$ and for decuplet baryons $J^P=(3/2)^+$, the Cabibbo angle can then be written as a function of the spin parity $J^P$ and deformation parameter $q$\footnote{The author is indebted to the referee for pointing out that formula (\ref{cabibbo}) with $\theta_{\mathbf{10}}=\frac{\pi}{14}$ can likewise be written in similar form by using, instead of $J^P$, the coefficient $(2J^P+1)/4$.}
\begin{eqnarray}\label{qcabibbo}
\theta_C=-iJ^P\ln{q}.
\end{eqnarray}  

\section{Discussion and conclusion}

In earlier works, the quantum group $SU_q(3)$ has been taken as a flavor symmetry and mass formulas of improved accuracy derived for octet and decuplet baryons. In the present paper we have built on these earlier works by taking into account (via Morpurgo's parametrization scheme) the electromagnetic contributions to octet and decuplet baryons mass. 

The deformed mass relations derived in \cite{gavrilik1998quantum,gavrilik2001quantum,gavrilik1997quantum} use the average of the isospin multiplet masses. The mass differences within the isoplets should not be ignored at the level of accuracy afforded by these deformed mass relations because they exceed the error in these formulas (for the best fit value of the deformation parameter $q$).

Accounting for the electromagnetic contributions to mass by selecting charge specific baryon masses (in a way that the electromagnetic mass contributions are balanced) rather than using isoplet mass averages leads to new octet and decuplet baryon mass relations accurate to 0.02\% and 0.08\% respectively. This represents more than a factor of 20 reduction in error compared to the standard GMO formulas.

A fixed value for the deformation parameter $q$ can be determined from the experimental data. The choice of $q=e^{i\pi/7}$ for octet baryons and $q=e^{i\pi/21}$ for decuplet baryons leads to a new relation between octet and decuplet baryons with an error of around 1.2\%, a slight improvement over an earlier formula (see equation (\ref{defoctdec}). The choice of $n=21$ for decuplet baryons suggests a simple formula for the Cabibbo angle in terms of $q$ and spin parity $J^P$. As a result, $q$ may not be a free parameter but rather be fixed from the baryons under consideration (either $J^P=(1/2)^+$ or $J^P=(3/2)^+$).

The electromagnetic contributions are only considered up to zeroth order. This is sufficient for our purposes as the resulting baryon mass relations are accurate to well within the experimental uncertainty. Once more accurate experimental data becomes available it will be worthwhile including higher order contributions.

Quantum groups are the result of one-parameter deformation in the universal enveloping algebra. These are just one type of algebraic deformation that can be considered \cite{gerstenhaber1964,nijenhuis1967}. Lie-type deformations (those that deform a Lie algebra) have proven very useful in generalizing spacetime symmetries \cite{mendes1994,ahluwalia2008ppa,Chryssomalakos2004,gresnigt2007sph}. $q$-Deformation on the other
 hand seem to have particular applications in generalized descriptions of internal and gauge symmetries \cite{Finkelstein2012,Finkelstein2005,finkelstein2007elementary}. Considered together therefore, it seems that Lie-type and $q$-deformations offer a consistent framework within which to develop physics in the 21st century \cite{gresnigt2015electroweak,Sternheimer2014,Flato1982,bonneau2003topological}.

\section*{Acknowledgments}
This work is supported in part by XJTLU research grant RDF-14-03-13 and Natural Science Foundation of China grant RR0116.

\bibliography{NielsReferences}  
\bibliographystyle{unsrt}  

\end{document}